\documentclass[aps,prl,showpacs,preprintnumbers,twocolumn]{revtex4-1}

\usepackage{amsmath,amssymb}
\usepackage{graphicx}
\usepackage{natbib}

\newcommand{\bnabla}{\boldsymbol{\nabla}\:}

\begin{document}

\title{Thermoelectricity in Confined Liquid Electrolytes}

\author{Mathias Dietzel} \email{dietzel@csi.tu-darmstadt.de} \author{Steffen Hardt}
\affiliation{Institute for Nano- and Microfluidics, Center of Smart Interfaces, TU Darmstadt\\
Alarich-Weiss-Str. 10, D-64287 Darmstadt, Germany}

\date{\today}

\begin{abstract}
The electric field in an extended phase of a liquid electrolyte exposed 
to a temperature gradient is attributed to different thermophoretic mobilities of the 
ion species. As shown herein, such Soret-type ion thermodiffusion is not required 
to induce thermoelectricity even in the simplest electrolyte if it is confined between 
charged walls. The space charge of the electric double layer leads to selective ion diffusion 
driven by a temperature-dependent electrophoretic ion mobility, which -for narrow channels- 
may cause thermo-voltages larger in magnitude than for the classical Soret equilibrium. \\ \\
This is the post-print authors' version of the manuscript, which was published in 
Physical Review Letters. doi:10.1103/PhysRevLett.116.225901  \copyright \ American Physical Society 2016.
\end{abstract}



\maketitle

The Seebeck effect describes the generation of a thermoelectric potential 
when a conductor is exposed to a temperature gradient $\bnabla T$ \cite{Hudak:JAP2008}. 
Thermoelectricity and its related effects are the cornerstones of key technologies for 
temperature measurements \cite{vanHerwaarden_SensAct1986}, refrigeration and recovery of waste heat 
\cite{Riffat:ApplThermEng2003,Bell:Science2008,Shakouri:AnnRevMatRes2011}, and has gained 
renewed interest within the realm of nanoscale transport processes \cite{Cahill:JAP2003}. 
While the charge carriers in the conduction band of semiconductors may generate a 
thermoelectric voltage without exhibiting a thermophoretic mobility 
\cite{Goldsmid:Thermoelectricity2010}, thermoelectricity in an extended (i.e. electroneutral) 
phase of a liquid electrolyte is based on thermophoresis of 
the dissolved ions species $k$ \cite{Guthrie:JChemPhys1949}. Their number concentrations 
$n_k$ align with $\bnabla T$ such that the ion fluxes driven by Fickian diffusion and 
thermophoresis balance each other. The overall salt concentration $n$ is then given by 
\cite{Wuerger:RepProgPhys2010}
\begin{equation}
\label{Eq:Soret-equilibrium} \frac{\bnabla n}{n} = -\Pi \frac{\bnabla T}{T}.
\end{equation}
Herein, to highlight the key effects, the discussion is focused on symmetric 
electrolytes of valence $\nu$ ($k=+$ for the cation and $k=-$ for the anion), for which 
electroneutrality implies $n_+ = n_- = n$. Consequently, the effective Soret 
coefficient simplifies to read $\Pi = (Q_+ + Q_-)/(2 k_{\textrm{B}} T)$. 
The Boltzmann constant is denoted by $k_\textrm{B}$, and the thermophoretic behavior 
of the ions is quantified in terms of the heats of transport, $Q_k$. For such 
simple electrolytes, the thermocell electric field is given by \cite{Wuerger:RepProgPhys2010}
\begin{equation}
\label{Eq:thermoelectric_pot} \boldsymbol{E}^{(\infty)}_Q = 
\frac{\Delta Q}{2 e \nu} \frac{\bnabla T}{T},
\end{equation}
i.e. it is solely generated by the difference in the thermophoretic mobilities of the ions 
expressed by $\Delta Q = Q_+ - Q_-$. The elementary charge is denoted by $e$. 
Equations (\ref{Eq:Soret-equilibrium}) and (\ref{Eq:thermoelectric_pot}) define the 
classical Soret equilibrium derived under the assumptions of vanishing flux densities, 
the absence of an advective velocity $u$ and electroneutrality throughout the domain.

\begin{figure}
	\centerline{\includegraphics[width=8.5cm]{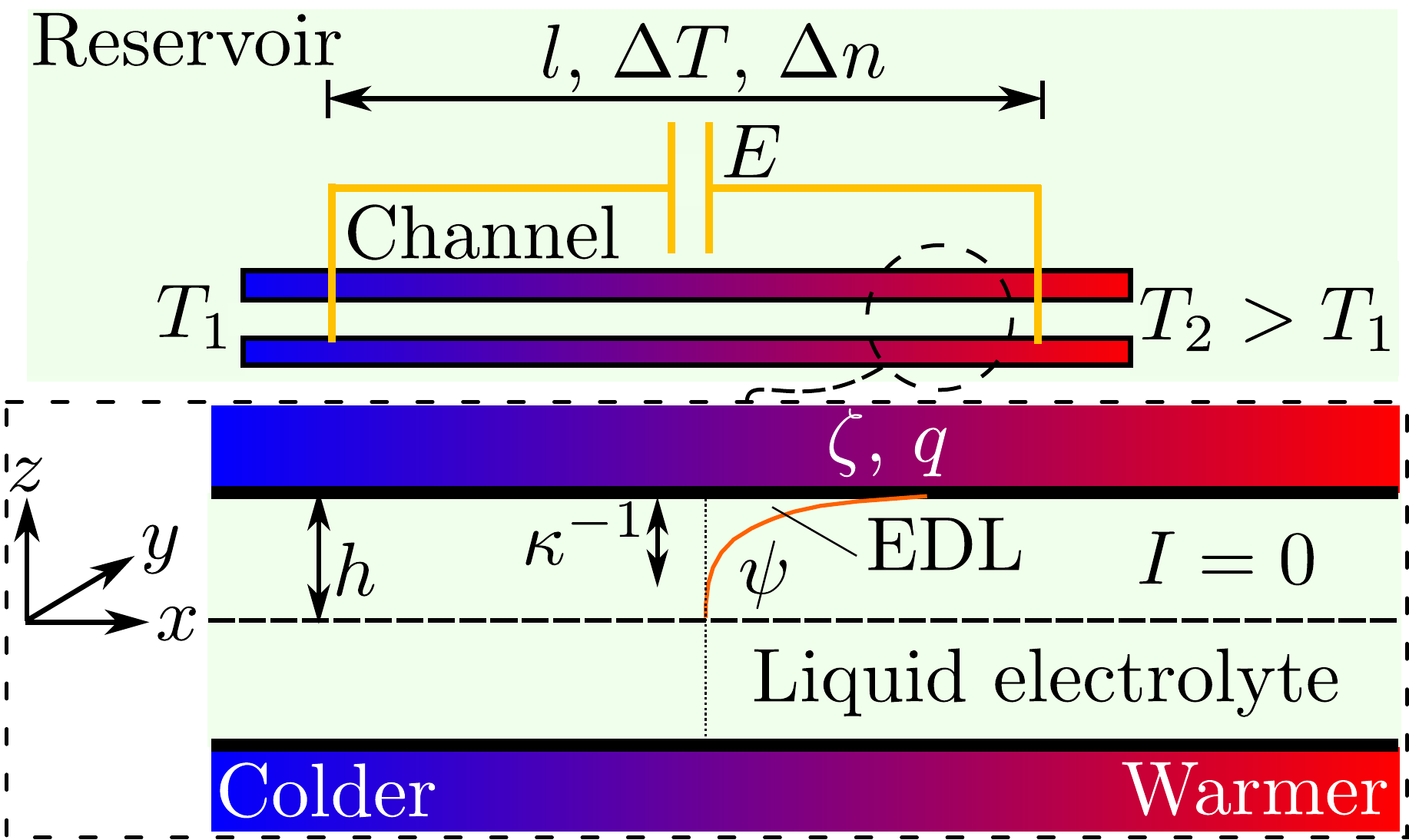}}
	\caption{Sketch of a slit channel of half-width $h$, submerged in an 
	extended, non-isothermal phase of an aqueous electrolyte. A temperature	difference 
	$\Delta T \leq T_2 - T_1$ is present at the channel walls over a length $l$, 
	leading to an induced thermoelectric field, $E$, and to a gradient in salt concentration, 
	$\Delta n/l$. The channel wall carries an electric surface charge density $q$ or 
	is kept at a constant $\zeta$-potential. The wall charge is screened by ions 
	in the electric double layer (EDL) of thickness $\kappa^{-1}$ with an internal 
	potential $\psi$. The total electric current, $I$, over the channel cross section vanishes.}
	\label{Fig:setup_schematic}
\end{figure}

Electroneutrality holds only for the bulk phase, where the influence of 
wall charges is negligibly small. Nevertheless, the 
relatively few theoretical investigations of the thermal membrane potential 
of electrolytes in charged pores commonly rely on the phenomenological 
theory of non-equilibrium thermodynamics and averaged transport numbers, 
neither explicitly resolving the ion distribution inside the pore, nor 
specifying the surface charge density, nor the pore size 
\cite{Hill:Nature1957,Tasaka:JChemPhys1965,Tasaka:PureApplChem1986,Gaeta:JPhysChem1992}. 
While numerous works discuss isothermal transport processes in charged nanochannels 
\cite{Schoch:RevModPhys2008}, practically no investigations are available addressing 
these interwoven issues in detail if the temperature is not constant. Electrolyte-filled 
nanopores and nanochannels with a temperature gradient play a key role for various phenomena. For 
instance, they are essential for the mechanisms by which organisms use ion channels to 
sense temperature \cite{McKemy:EurJPhysiol2007,Reid:Nature2001,Viana:NatureNeuro2002}. 
In addition, the non-isothermal ion transport in porous membranes is a 
promising candidate for the development of efficient thermoelectric energy conversion 
techniques \cite{Sandbakk:JMembrScie2013,Xu:NanoEnergy2012}. The purpose of this Letter 
is to analyze the non-advective transport phenomena occurring in a slit nanochannel 
with charged walls and filled with an electrolyte under application of a temperature gradient 
along the channel. As will be shown, the presence of an electric double layer (EDL) alters 
the Soret voltage given by (\ref{Eq:thermoelectric_pot}), but also induces an additional 
thermoelectric voltage due to the temperature-dependent electrophoretic ion mobility alone, 
without relying on the intrinsic Soret effect quantified by the parameters $Q_k$. To the best of 
our knowledge, such a mechanism has never been explicitly described before. Thermoosmotic 
effects arising from the mechanical imbalance of the non-isothermal ion cloud of the EDL 
will be disregarded, i.e. the momentum equations are not solved for. In preliminary, 
so far unpublished work, they were found to be weak compared to the phenomena discussed 
herein. 

Figure \ref{Fig:setup_schematic} depicts a schematic of the investigated system. 
The wall temperature of a long slit channel of half-width $h$ filled with a dilute 
electrolyte uniformly increases by $\Delta T$ over a length of $l$. The analysis of 
this paper is based on the leading-order contribution of an asymptotic 
expansion in the small parameter $A=h/l$, while the dimensionless axial gradients of any 
quantity $n_k, T, ...$ are assumed to be small. As verified in the Supplemental Material, 
advection, viscous dissipation and Joule heating can be neglected in the 
energy equation \cite{SM}$^{(\S 1)}$. Hence, to first order in $A$, the temperature gradient in the 
interior of the channel is given by $\bnabla T = (\Delta T/l,0)$, even if the thermal 
conductivity of the electrolyte varies with temperature. With $D_k$ being 
the Fickian diffusion coefficients of the ion species, the ionic P\'eclet numbers, 
$Pe_k = u l/D_k$, are negligibly small in the present system \cite{SM}$^{(\S 1)}$. 
The Nernst-Planck equations (NPE), governing the ion concentrations, simplify to 
$\bnabla \cdot \boldsymbol{j}_k = 0$ \cite{SM}$^{(\S 2)}$, with
\begin{equation}
\label{Eq:ion_flux_density} -\boldsymbol{j}_k = D_k\bnabla n_k + n_k \mu_k\bnabla{T} + e\nu_k n_k \omega_k\bnabla \phi
\end{equation}
being the ion flux densities, where $\mu_k \equiv D_k Q_k/(k_\textrm{B} T^2)$ 
and $\boldsymbol{j}_k = (j_{k,x},j_{k,z})$. The overall electric field 
$\bnabla \phi = \bnabla \psi- \boldsymbol{E}$ is the sum of the EDL field $\bnabla \psi$, 
fulfilling the Poisson equation, and an induced electric field, $\boldsymbol{E}$, with 
vanishing associated charge density (source-free) \cite{Fair:JChemPhys1971}. For $A^2 
\ll 1$, the Laplace equation and the symmetry condition at the channel center imply that 
$\boldsymbol{E} \approx (E,0)$ \cite{SM}$^{(\S 3)}$. The electrophoretic ion mobilities are given by 
the classical Stokes-Einstein-relation $\omega_k = D_k/(k_{\textrm{B}} T)$. 
To leading order in $A$ of the NPE and again incorporating the symmetry at the 
channel center, the ion concentrations are determined from (\ref{Eq:ion_flux_density}) 
by $j_{k,z} = 0$. Together with $\partial_z T \equiv \partial T/\partial z = 0$, 
one finds that the local ion number concentrations are given by \cite{SM}$^{(\S 3)}$
\begin{equation} 
\label{Eq:Boltzmann_dim} n_k = n_{k,0} \textrm{exp} \left(-\frac{e \nu_k \psi}{k_{\textrm{B}} T}\right),
\end{equation}
which resemble the Boltzmann-distribution. The local ion concentrations at 
$\psi=0$ (electroneutral region) are denoted by $n_{k,0}$, which may be a function 
of $x$ \cite{Sasidhar:JCollIntScie1982}. For a symmetric electrolyte, 
$n_{k,0} \equiv n$ for each ion species. Despite its 
familiar appearance, equation (\ref{Eq:Boltzmann_dim}) is a consequence of the 
smallness of $A$ and the symmetry along the channel center, rather than of directly 
imposing thermodynamic equilibrium. Unlike in the conventional Soret equilibrium, 
$n_+ \neq n_- \neq n$. By inserting (\ref{Eq:Boltzmann_dim}) in 
(\ref{Eq:ion_flux_density}), the axial flux densities are given by
\begin{equation} 
\label{Eq:ion_flux_dens_ax} -\frac{j_{k,x}}{n_k D_k} = d_x \textrm{ln}(n) 
+ \frac{e \nu_k}{k_{\textrm{B}} T}\left[-E  + \left(\frac{Q_k}{e \nu_k} + \psi \right) d_x \textrm{ln}(T)\right],
\end{equation}
where $d_x(.) \equiv d(.)/dx$. For simple salts, the coefficients $D_k$ are 
very similar to each other. Focusing on the essential effects, identical 
$D_k \equiv D$ are assumed in the following. However, $D$ does not 
need to be a constant and may vary with $T$. Under no external electric load, 
$E$ is calculated by setting the overall electric current, 
$I = e \nu \int^h_0{(j_{+,x} - j_{-,x})dz}$, to zero. This is equivalent to what is 
done in studies of thermoelectricity in bulk electrolytes \cite{Wuerger:RepProgPhys2010}. 
For the Seebeck coefficient $S \equiv E/d_x T$ one finds \cite{SM}$^{(\S 4)}$
\begin{equation} 
\label{Eq:Thermoelectricity_confined} S = S_Q + S_\psi,
\end{equation}
where
\begin{equation} 
\label{Eq:Thermoelectricity_confined_Soret} S_Q = 
\frac{1}{T}\frac{\Delta Q}{2 e \nu} \frac{\int^h_0{\textrm{e}^{-\Psi}dz} 
- \frac{q}{2 e \nu n}}{\int^h_0{\textrm{cosh}(\Psi)dz}},
\end{equation}
\begin{equation} 
\label{Eq:Thermoelectricity_confined_Tdep} S_\psi = 
\frac{1}{T}\frac{\int^h_0{\psi \textrm{cosh}(\Psi)dz}}{\int^h_0{\textrm{cosh}(\Psi)dz}}. 
\end{equation}
The surface charge density is denoted by $q = - \epsilon (\partial_z \psi)_{|z=h}$, 
with $\epsilon$ being the (temperature-dependent) dielectric permittivity, while 
$\Psi \equiv e \nu \psi/(k_{\textrm{B}} T)$. To derive $E$ and since the functional 
form of $n$ cannot be determined within the employed approximation scheme, 
$d_x \textrm{ln}(n) = d_x n/n$ was expressed by (\ref{Eq:Soret-equilibrium}). While being 
accurate for channels with non-overlapping EDLs, for narrower channels this assumption can be 
justified by viewing the channel as being submerged in a large, non-isothermal tank and 
referring to a system in electrochemical equilibrium for every local value of 
$T$ \cite{SM}$^{(\S 4)}$. Furthermore, neglecting terms of ${\cal O}(A^2)$, the Poisson equation 
$\bnabla \cdot (\epsilon \bnabla \psi) = -\rho_f$ was used, with $\rho_f = e \nu (n_+ - n_-)$ 
being the charge density. Note that, to first order in $A$, the term $\bnabla \epsilon \cdot \bnabla \psi$ 
can be neglected in the Poisson equation, even though $\epsilon = \epsilon(T)$.

On the one hand, $S_Q$ defined by (\ref{Eq:Thermoelectricity_confined_Soret}) 
expresses the thermoelectric field caused by the Soret-type thermophoretic ion motion 
under confinement. The presence of an EDL modifies its corresponding bulk value, given in 
form of the classical Soret equilibrium by (\ref{Eq:thermoelectric_pot}), to which 
$S d_x T$ reduces for an uncharged or very wide channel. On the other hand, 
$S_Q$ vanishes even under confinement if the heats of transport of both ion species 
are identical ($\Delta Q = 0$). However, in that case the overall thermoelectric field 
does not necessarily vanish but is given by $(S)_{|\Delta Q \equiv 0} = S_\psi$ 
alone. If $\Delta Q \neq 0$, $S_Q$ and $S_\psi$ are additive. 
Since advection is completely neglected herein, the latter field does not have a thermoosmotic 
origin \cite{SM}$^{(\S 5)}$. Instead, according to (\ref{Eq:Boltzmann_dim}), the temperature-dependent electrophoretic 
ion mobility implies axial gradients of $n_k$ within the EDL, which are additive to 
(\ref{Eq:Soret-equilibrium}), while the magnitude of the affiliated (Fickian) diffusion fluxes 
depend on the polarity of the ion species. This gives rise to charge separation and induces an electric field. 
To the best of our knowledge, despite being a direct consequence of the fundamental 
Stokes-Einstein-equation, such an effect has never been described before.

To further evaluate $S$, $\psi$ has to be determined by solving the Poisson equation. 
Along with equation (\ref{Eq:Boltzmann_dim}), one finds to leading order in $A$
\begin{equation}
\label{Eq:Poisson_1D} \partial^2_z \Psi \approx \kappa^2 \textrm{sinh} (\Psi).
\end{equation}
The local Debye parameter is given by 
$\kappa = \sqrt{2 e^2 \nu^2 n/(\epsilon k_{\textrm{B}} T)}$. Note that since 
$T=T(x)$, $\Psi = \Psi(x)$ and $\kappa = \kappa(x)$ as well 
(we set $\kappa_\textrm{r} = (\kappa)_{|x=0}$, with $T=T_{\textrm{r}}$ 
and $n = n_\textrm{r}$ as a reference). However, the modification of $S_\psi$ 
caused by these dependences is of higher order in $d_x T$ and can be neglected in most situations. 
The solution of (\ref{Eq:Poisson_1D}) is formally identical to the one of the isothermal 
Poisson-Boltzmann (PB) equation. Within the Debye-H\"uckel (DH) approximation ($|\Psi|<1$) it reads 
$\psi^{(\textrm{DH})} = \zeta \textrm{cosh}(\kappa z)/\textrm{cosh}(\overline{\kappa})$, 
where $\overline{\kappa} = \kappa h$, and $\zeta$ is the $\zeta$-potential at the 
slipping plane of the wall. With this, the integrals in (\ref{Eq:Thermoelectricity_confined_Soret}) 
and (\ref{Eq:Thermoelectricity_confined_Tdep}) can be evaluated, yielding
\begin{equation} 
\label{Eq:Thermoelectricity_confined_Soret_DH} S^{(\textrm{DH})}_Q = 
\frac{1}{T}\frac{\Delta Q}{2 e \nu}
\frac{1}{1+\frac{\overline{\zeta}^2}{4}\left[\frac{\textrm{tanh}(\overline{\kappa})}{\overline{\kappa}}\!+\!\frac{1}{\textrm{cosh}^2(\overline{\kappa})}\right]},
\end{equation}
and
\begin{equation} 
\label{Eq:Thermoelectricity_confined_Tdep_DH} S^{(\textrm{DH})}_{\psi} =
\frac{\zeta}{T} \frac{\textrm{tanh}(\overline{\kappa})}{\overline{\kappa}} 
\frac{1+\frac{\overline{\zeta}^2}{2}\left[\frac{\textrm{tanh}^2(\overline{\kappa})}{3}\!+\!\frac{1}{\textrm{cosh}^2(\overline{\kappa})}\right]}
{1+\frac{\overline{\zeta}^2}{4}\left[\frac{\textrm{tanh}(\overline{\kappa})}{\overline{\kappa}}\!+\!\frac{1}{\textrm{cosh}^2(\overline{\kappa})}\right]},
\end{equation}
with $\overline{\zeta} \equiv e \nu \zeta/(k_{\textrm{B}} T)$. 
According to (\ref{Eq:Thermoelectricity_confined_Soret_DH}) and for sufficiently small 
$\zeta$, the effect of the EDL on the Soret voltage is negligibly small. 
At constant $\zeta$, for $\overline{\kappa} \rightarrow 0$ one has 
$S^{(\textrm{DH})}_{\psi} \rightarrow \zeta/T$, 
whereas $S^{(\textrm{DH})}_{\psi}$ vanishes for $\overline{\kappa} \rightarrow \infty$. 
Hence, the thermoelectric field induced by a temperature-dependent electrophoretic 
ion mobility is a confinement effect and dominant in charged, narrow channels. 

\begin{figure}
	\centerline{\includegraphics[width=8.5cm]{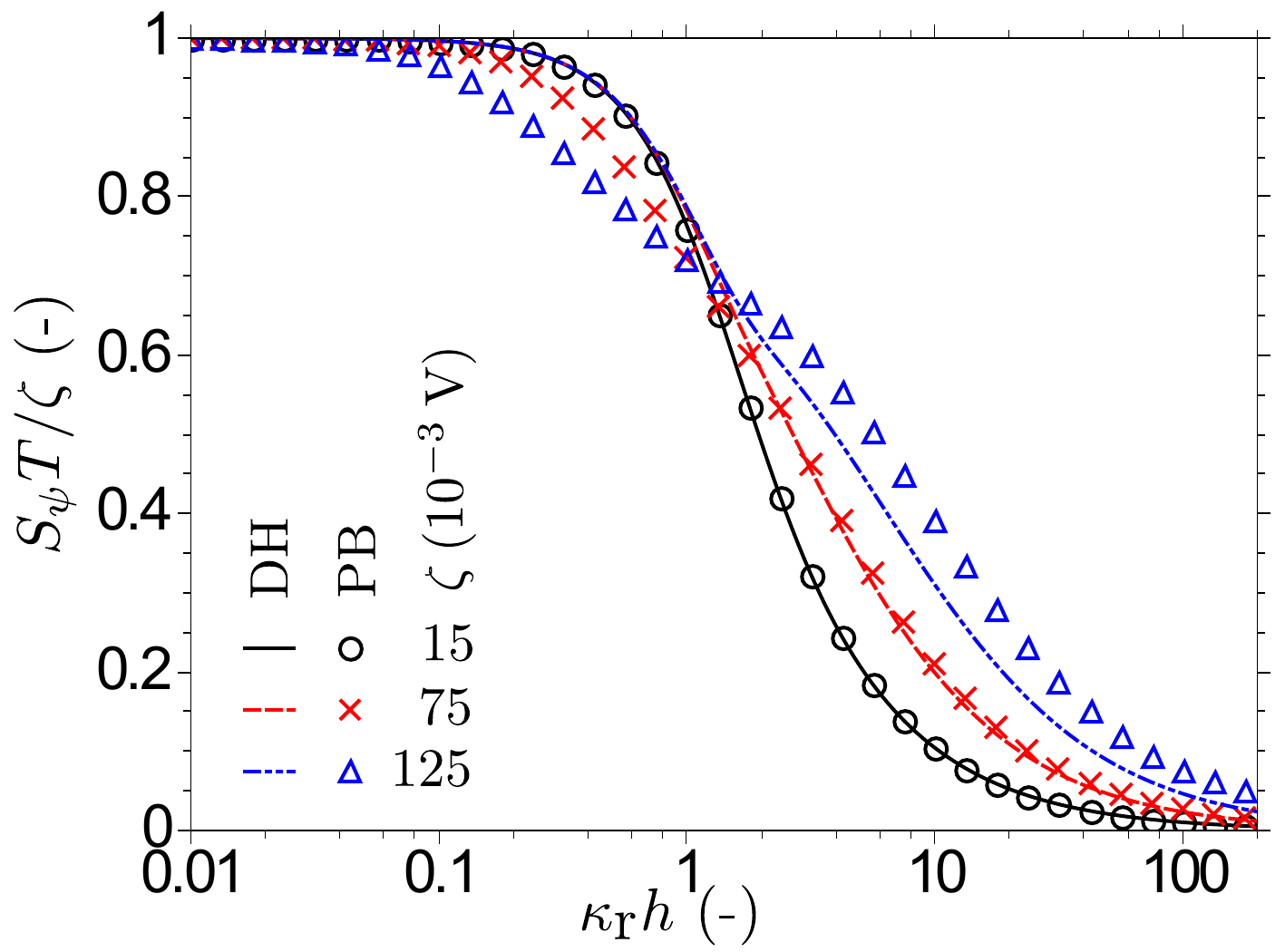}}
	\caption{Seebeck coefficient $S_\psi$ relative to $\zeta/T$ and as 
	a function of the nominal Debye parameter 
	$\overline{\kappa}_\textrm{r} = \kappa_\textrm{r} h$. Results based on 
	the Debye-H\"uckel (DH) approximation (lines without symbols, computed by 
	(\ref{Eq:Thermoelectricity_confined_Tdep_DH})), 
	are compared to solutions based on a numerical evaluation (PB, Poisson-Boltzmann) of 
	(\ref{Eq:Poisson_1D}) (lines with symbols). All data were calculated for $T = 298\:\textrm{K}$.}
	\label{Fig:thermopot_zetavary}
\end{figure}

In figure \ref{Fig:thermopot_zetavary}, the Seebeck coefficient $S_\psi$, 
non-dimensionalized by $\zeta/T$, is plotted as a function of 
$\overline{\kappa}_\textrm{r}$, while $\zeta = [15,75,125] \cdot 10^{-3}\:\textrm{V}$. 
The solutions according to the DH-approximation are compared to those based on numerical 
evaluations of (\ref{Eq:Poisson_1D}) \cite{SM}$^{(\S 6)}$. With increasing $\overline{\kappa}_\textrm{r}$, 
all curves continuously decrease from unity to zero. The PB-solutions and the DH-approximation 
are indistinguishable for $\zeta = 15\cdot 10^{-3}\:\textrm{V}$ and almost identical for
$\zeta = 75\cdot 10^{-3}\:\textrm{V}$ if $\overline{\kappa}_\textrm{r} > 2$. If 
$\overline{\kappa}_\textrm{r} \lesssim 2$ while $\zeta = 75\cdot 10^{-3}\:\textrm{V}$, 
the DH-approximation overpredicts $S_\psi$. This occurs also for 
$\zeta = 125\cdot 10^{-3}\:\textrm{V}$, whereas in that case for 
$\overline{\kappa}_\textrm{r} \gtrsim 2$ $S_\psi$ is underpredicted. All data shown 
were evaluated at $T = T_{\textrm{r}} = 298\:\textrm{K}$. Corresponding 
calculations at $T = 308\:\textrm{K}$ (using $\Pi/T = 5 \cdot 10^{-3}\:\textrm{K}^{-1}$ 
\cite{Snowdon:TransFaradaySoc1960II,Agar:ProcRsSocLondA1960,Snowdon:TransFaradaySoc1960} and 
$d_T \epsilon/\epsilon = -5.1 \cdot 10^{-3}\:\textrm{K}$ \cite{Buchner:PhysChemA1999}) 
to estimate the magnitude of possible non-linear effects due to 
$\overline{\kappa} = \overline{\kappa}(T)$ gave practically indistinguishable results (not shown).

\begin{figure}
	\centerline{\includegraphics[width=8.5cm]{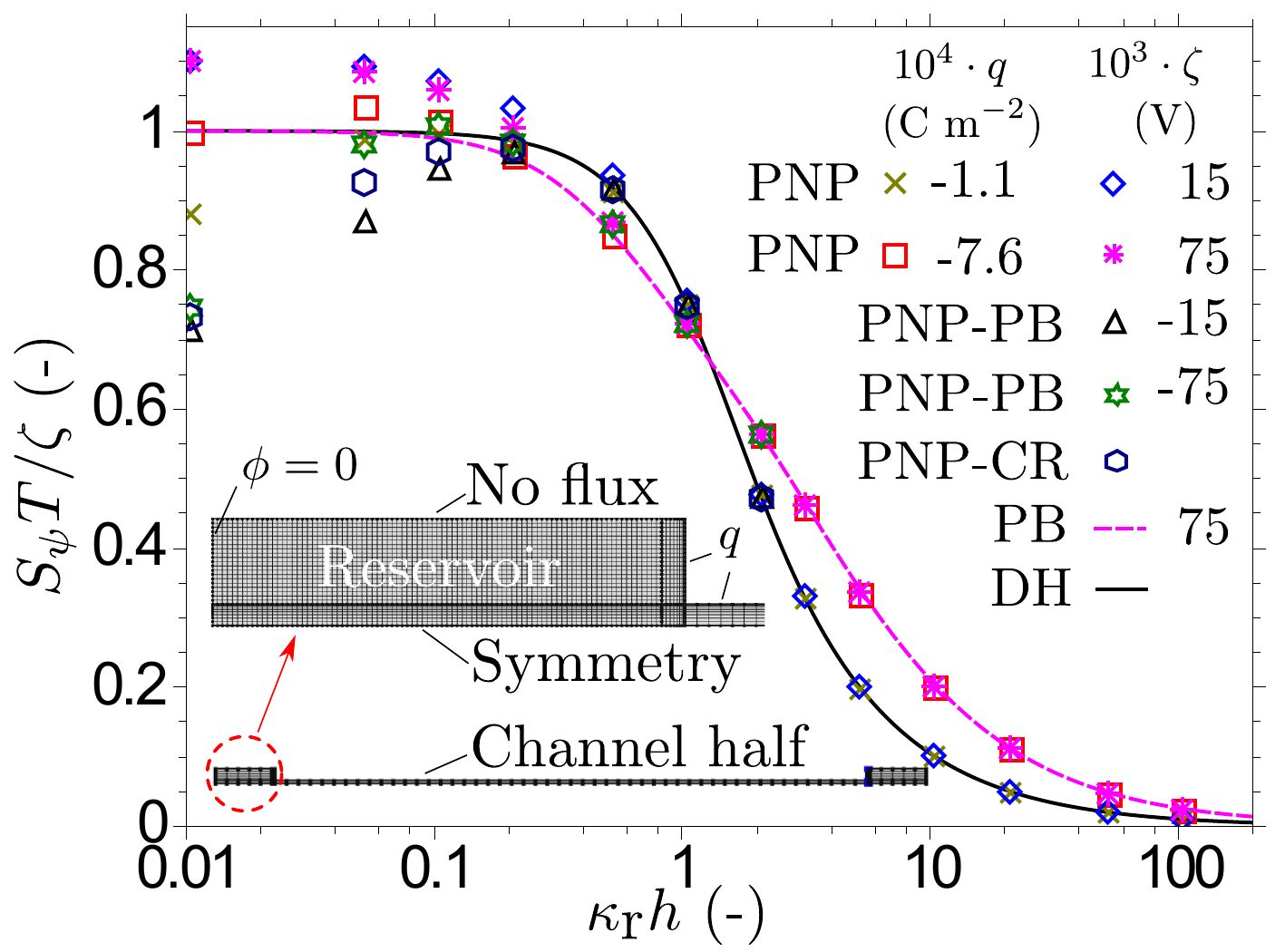}}
	\caption{Seebeck coefficient $S_\psi$ relative to $\zeta/T$ and as a function of the 
	nominal Debye parameter 
	$\overline{\kappa}_\textrm{r} = \kappa_\textrm{r} h$. The data points are obtained from a full 
	numerical simulation of the Poisson equation and the Nernst-Planck equation (PNP), 
	without relying on (\ref{Eq:Boltzmann_dim}). Either the surface charge density $q$ 
	or the $\zeta$-potential is held constant. The PNP-model is compared with the Debye-H\"uckel 
	(DH) approximation (\ref{Eq:Thermoelectricity_confined_Tdep_DH}) as well as with the 
	Poisson-Boltzmann (PB) model, which is based on a numerical evaluation of (\ref{Eq:Poisson_1D}). 
	For selected cases (PNP-PB), $q$ imposed in the PNP-simulation 
	as a boundary condition	was pre-determined for a given $\overline{\kappa}_\textrm{r}$ from an 
	analytical solution of (\ref{Eq:Poisson_1D}), where $\zeta$ was set either 
	to $-15$ or $-75\cdot10^{-3}\:\textrm{V}$. For a $\textrm{pH}$-value of $4$, the 
	PNP-simulation were also combined with a temperature-dependent charge regulation model 
	(PNP-CR), which is detailed in the Supplemental Material \cite{SM}$^{(\S 9)}$.The temperature 
	was set to $T=298\:\textrm{K}$.}
	\label{Fig:thermopot_fullnum}
\end{figure}

The NPE treat the ions as point charges, so that the effects of the finite ion size 
\cite{Tessier:Electrophoresis2006} and ion-ion-correlations on steric and Coulombic 
interactions \cite{Bazant:AdvCollIntScie2009} are neglected. This is permissible for 
ion concentrations and $\zeta$-potentials not significantly exceeding 
$n_\textrm{r} = 0.01\:\textrm{M}$ and 
$\zeta = 125 \cdot 10^{-3}\:\textrm{V}$ \cite{Cervera:MicroNanofluidics2010}. 
The effect under study is at its maximum for $\overline{\kappa} \rightarrow 0$. 
In this limit the ion cloud does not completely screen the surface charge, i.e. $\psi$ is 
non-vanishing at the center of the channel. While for sufficiently wide channels the Gouy-Chapman (GC) 
equation implies the equivalency of a constant $\zeta$-potential and a constant value of $q$ 
\cite{Russel:CambrUnivPress1989}, this does not hold for channels with pronounced EDL-overlap 
\cite{Movahed:Electrophoresis2011}. In this case, the variation of the electrokinetic 
characteristics of a system as function of $\overline{\kappa}_\textrm{r}$ may be different depending 
on whether a constant value of $\zeta$ or of $q$ is imposed. 
Furthermore, the constituting equations of the PB-model are derived in the 
framework of a first-order expansion in $A$. Especially the validity of (\ref{Eq:Boltzmann_dim}) 
in case of overlapping EDLs has to be confirmed, since the reference concentrations at $\psi=0$, 
$n_{k,0}$, can no longer refer to a location inside the channel. Such issues can be avoided 
by a (numerically solved) model solely based on the coupled Poisson- and Nernst-Planck (PNP) 
equations, without relying on (\ref{Eq:Boltzmann_dim}). In figure \ref{Fig:thermopot_fullnum}, 
the results of such simulations for the given system are shown. The computations \cite{SM}$^{(\S 7)}$ 
followed the basic strategy outlined in refs. 
\cite{Daiguji:ElectrochemComm2006,Movahed:Electrophoresis2011} 
and were conducted using Comsol Multiphysics 4.3a \cite{Comsol2014}. While varying 
$\overline{\kappa}_\textrm{r}$, either a constant $\zeta$-potential 
($\zeta =[15,75]\cdot 10^{-3} \:\textrm{V}$) or a constant surface 
charge density ($q = [-1.1,-7.6] \cdot 10^{-4}\:\textrm{C m}^{-2}$) 
was imposed along the channel walls, where the mapping between $\zeta$ 
and $q$ is provided by the GC-model \cite{SM}$^{(\S 7)}$. Given the negligible extent 
of advection, the Navier-Stokes equations were not included in the model. 

In figure \ref{Fig:thermopot_fullnum} the (relative) Seebeck coefficient $S_\psi T/\zeta$ is shown as a function 
of $\overline{\kappa}_\textrm{r}$. For the numerical simulations (depicted as symbols), the 
variation of $\overline{\kappa}_\textrm{r}$ was achieved by a variation of $h$, with the nominal 
EDL-thickness being held constant at $\kappa^{-1}_\textrm{r} \approx 10^{-7}\:\textrm{m}$. For 
low $\zeta$ or $q$, the numerical results are compared with 
the DH-approximation, while for more strongly charged walls they are compared with the PB-model. 
For $\overline{\kappa}_\textrm{r} > 0.5$, the PNP-solutions fully agree with the corresponding 
(quasi-) analytical solutions (DH or PB). From the PNP-simulations conducted at constant 
$q$, it follows that the surface charge is almost completely screened if $h$ 
is at least twice the nominal EDL-thickness. 

The PNP-simulations of the cases with overlapping EDLs ($0.01 \leq \overline{\kappa}_\textrm{r} \leq 2$) 
were repeated by imposing $q$ at the charged walls, with its value -for 
given $\zeta$ and $\overline{\kappa}_\textrm{r}$- being individually pre-calculated 
by the analytical solution of the PB-equation (\ref{Eq:Poisson_1D}) 
\cite{Behrens:PRE1999} rather than using the GC-model \cite{SM}$^{(\S 8)}$. For $0.2 \leq \overline{\kappa}_\textrm{r} \leq 2$ 
at $\zeta = -15$ or $-75 \cdot 10^{-3}\:\textrm{V}$, the corresponding 
results (see figure \ref{Fig:thermopot_fullnum}) agree well with those where a 
constant $\zeta$-potential is imposed directly. For decreasing $\overline{\kappa}_\textrm{r}$ below 
$0.2$, the PNP-model increasingly deviates from the DH- and the PB-model for 
any considered $\zeta$-potential or $q$. This is likely caused by an 
insufficient length of the channel in the computational domain, but could not be resolved 
with the available computational resources. From the PNP-simulations it was also found that 
the Soret voltage under confinement, expressed by (\ref{Eq:Thermoelectricity_confined_Soret}), 
deviates from its bulk value $S^{(\infty)}_Q$ by not more than $1\%$ for any 
channel width.

The invariance of either the $\zeta$-potential or $q$ along the 
channel wall does not necessarily hold for a non-isothermal channel, and both 
parameters might be a function of temperature \cite{Reppert:JGeoPhysRes2003}. 
Such questions can be addressed by detailing the surface charge formation process 
\cite{Behrens:JChemPhys2001,Revil:PhysRevB1997}. For a silica channel with its surface 
charge being mainly formed by the dissociation of silanol-groups, corresponding 
results are included in figure \ref{Fig:thermopot_fullnum}. For a $\textrm{pH}$-value of $4$, 
the model was calibrated with experimental data available for $\textrm{NaCl}$ as electrolyte 
\cite{SM}$^{(\S 9)}$. It is apparent that the results using the (temperature-dependent) charge regulation 
model follow closely those of the other approaches. Consequently, the thermoelectricity 
described by (8) (and in the DH-limit by (11)), being the main result of this work, is 
remarkably robust for values of $\overline{\kappa}_\textrm{r}$ larger than about $0.2$.

The prediction of $S_\psi$ is firmly linked to the particular expression of the 
electrophoretic ion mobility in form of the Einstein-Smoluchowski equation, which 
is the simplest form of a fluctuation-dissipation relation under infinite dilution. 
It was derived under the assumption of isothermal conditions \cite{Einstein:AnnPhysik1905}, 
and its use in non-isothermal systems of low ionic strength is acceptable only if the 
momentum relaxation time of an ion is much smaller than the time the particle takes to experience 
a temperature change \cite{Stolovitzky:PhysicsLettA1998}. Herein, the ratio between 
these characteristic time scales is ${\cal O}(10^{-8})$ \cite{SM}$^{(\S 10)}$. For systems of higher 
ionic strength and complex electrolyte solutions, the electrophoretic ion mobility 
might be itself a non-trivial function of temperature \cite{Rogacs:AnalChem2013}, which is 
beyond the scope of the present work.

Our results indicate that the thermoelectric voltage of dilute electrolytes 
in confined geometry may be quite different from its bulk counterpart. 
While the presence of the latter is intrinsically linked to different thermophoretic 
mobilities of the ion species alone, the former may be present even if the heats of transport 
of each ion species are identical ($\Delta Q = 0$) or very small. In this case, the thermoelectric 
voltage under confinement is solely proportional to the $\zeta$-potential or surface charge density 
of the channel and reaches its maximum in the limit of $\overline{\kappa}_\textrm{r} \rightarrow 0$. 
For narrow, highly charged channels and within the validity range of the presented theory, such 
thermo-voltages might be up to $30$ times larger than the values of the conventional Soret voltage
being typical for simple monovalent electrolytes in the bulk \cite{SM}$^{(\S 11)}$. Our findings can be used as 
a novel method to determine the $\zeta$-potential of nanochannels and biological ion channels, 
while also being of interest for the design of novel small-scale heat exergy (i.e. availability) 
recovery devices. Even though the presented theory is strictly valid only for domains of small 
aspect ratio, the underlying mechanism may still have an impact on the thermophoretic motion of 
larger, charge-stabilized particles \cite{Braibanti:PRL2008,Majee:PRE2011} and biological 
molecules \cite{Reichl:PRL2014}.

\bibliography{references}

\begin{thebibliography}{55}%
\makeatletter
\providecommand \@ifxundefined [1]{%
 \@ifx{#1\undefined}
}%
\providecommand \@ifnum [1]{%
 \ifnum #1\expandafter \@firstoftwo
 \else \expandafter \@secondoftwo
 \fi
}%
\providecommand \@ifx [1]{%
 \ifx #1\expandafter \@firstoftwo
 \else \expandafter \@secondoftwo
 \fi
}%
\providecommand \natexlab [1]{#1}%
\providecommand \enquote  [1]{``#1''}%
\providecommand \bibnamefont  [1]{#1}%
\providecommand \bibfnamefont [1]{#1}%
\providecommand \citenamefont [1]{#1}%
\providecommand \href@noop [0]{\@secondoftwo}%
\providecommand \href [0]{\begingroup \@sanitize@url \@href}%
\providecommand \@href[1]{\@@startlink{#1}\@@href}%
\providecommand \@@href[1]{\endgroup#1\@@endlink}%
\providecommand \@sanitize@url [0]{\catcode `\\12\catcode `\$12\catcode
  `\&12\catcode `\#12\catcode `\^12\catcode `\_12\catcode `\%12\relax}%
\providecommand \@@startlink[1]{}%
\providecommand \@@endlink[0]{}%
\providecommand \url  [0]{\begingroup\@sanitize@url \@url }%
\providecommand \@url [1]{\endgroup\@href {#1}{\urlprefix }}%
\providecommand \urlprefix  [0]{URL }%
\providecommand \Eprint [0]{\href }%
\providecommand \doibase [0]{http://dx.doi.org/}%
\providecommand \selectlanguage [0]{\@gobble}%
\providecommand \bibinfo  [0]{\@secondoftwo}%
\providecommand \bibfield  [0]{\@secondoftwo}%
\providecommand \translation [1]{[#1]}%
\providecommand \BibitemOpen [0]{}%
\providecommand \bibitemStop [0]{}%
\providecommand \bibitemNoStop [0]{.\EOS\space}%
\providecommand \EOS [0]{\spacefactor3000\relax}%
\providecommand \BibitemShut  [1]{\csname bibitem#1\endcsname}%
\let\auto@bib@innerbib\@empty
\bibitem [{\citenamefont {Hudak}\ and\ \citenamefont
  {Amatucci}(2008)}]{Hudak:JAP2008}%
  \BibitemOpen
  \bibfield  {author} {\bibinfo {author} {\bibfnamefont {N.~S.}\ \bibnamefont
  {Hudak}}\ and\ \bibinfo {author} {\bibfnamefont {G.~G.}\ \bibnamefont
  {Amatucci}},\ }\href@noop {} {\bibfield  {journal} {\bibinfo  {journal}
  {J.~Appl.~Phys.}\ }\textbf {\bibinfo {volume} {103}},\ \bibinfo {pages}
  {101301} (\bibinfo {year} {2008})},\ \bibinfo {note} {and references
  therein}\BibitemShut {NoStop}%
\bibitem [{\citenamefont {van Herwaarden}\ and\ \citenamefont
  {Sarro}(1986)}]{vanHerwaarden_SensAct1986}%
  \BibitemOpen
  \bibfield  {author} {\bibinfo {author} {\bibfnamefont {A.~W.}\ \bibnamefont
  {van Herwaarden}}\ and\ \bibinfo {author} {\bibfnamefont {P.~M.}\
  \bibnamefont {Sarro}},\ }\href@noop {} {\bibfield  {journal} {\bibinfo
  {journal} {Sens.~Actuators}\ }\textbf {\bibinfo {volume} {10}},\ \bibinfo
  {pages} {321} (\bibinfo {year} {1986})}\BibitemShut {NoStop}%
\bibitem [{\citenamefont {Riffat}\ and\ \citenamefont
  {Ma}(2003)}]{Riffat:ApplThermEng2003}%
  \BibitemOpen
  \bibfield  {author} {\bibinfo {author} {\bibfnamefont {S.~B.}\ \bibnamefont
  {Riffat}}\ and\ \bibinfo {author} {\bibfnamefont {X.}~\bibnamefont {Ma}},\
  }\href@noop {} {\bibfield  {journal} {\bibinfo  {journal}
  {Appl.~Therm.~Eng.}\ }\textbf {\bibinfo {volume} {23}},\ \bibinfo {pages}
  {913} (\bibinfo {year} {2003})}\BibitemShut {NoStop}%
\bibitem [{\citenamefont {Bell}(2008)}]{Bell:Science2008}%
  \BibitemOpen
  \bibfield  {author} {\bibinfo {author} {\bibfnamefont {L.~E.}\ \bibnamefont
  {Bell}},\ }\href@noop {} {\bibfield  {journal} {\bibinfo  {journal}
  {Science}\ }\textbf {\bibinfo {volume} {321}},\ \bibinfo {pages} {1457}
  (\bibinfo {year} {2008})}\BibitemShut {NoStop}%
\bibitem [{\citenamefont {Shakouri}(2011)}]{Shakouri:AnnRevMatRes2011}%
  \BibitemOpen
  \bibfield  {author} {\bibinfo {author} {\bibfnamefont {A.}~\bibnamefont
  {Shakouri}},\ }\href@noop {} {\bibfield  {journal} {\bibinfo  {journal}
  {Annu.~Rev.~Mater.~Res.}\ }\textbf {\bibinfo {volume} {41}},\ \bibinfo
  {pages} {399} (\bibinfo {year} {2011})}\BibitemShut {NoStop}%
\bibitem [{\citenamefont {Cahill}\ \emph {et~al.}(2003)\citenamefont {Cahill},
  \citenamefont {Ford}, \citenamefont {Goodson}, \citenamefont {Mahan},
  \citenamefont {Majumdar}, \citenamefont {Maris}, \citenamefont {Merlin},\
  and\ \citenamefont {Phillpot}}]{Cahill:JAP2003}%
  \BibitemOpen
  \bibfield  {author} {\bibinfo {author} {\bibfnamefont {D.~G.}\ \bibnamefont
  {Cahill}}, \bibinfo {author} {\bibfnamefont {W.~K.}\ \bibnamefont {Ford}},
  \bibinfo {author} {\bibfnamefont {K.~E.}\ \bibnamefont {Goodson}}, \bibinfo
  {author} {\bibfnamefont {G.~D.}\ \bibnamefont {Mahan}}, \bibinfo {author}
  {\bibfnamefont {A.}~\bibnamefont {Majumdar}}, \bibinfo {author}
  {\bibfnamefont {H.~J.}\ \bibnamefont {Maris}}, \bibinfo {author}
  {\bibfnamefont {R.}~\bibnamefont {Merlin}}, \ and\ \bibinfo {author}
  {\bibfnamefont {S.~R.}\ \bibnamefont {Phillpot}},\ }\href@noop {} {\bibfield
  {journal} {\bibinfo  {journal} {J.~Appl.~Phys.}\ }\textbf {\bibinfo {volume}
  {93}},\ \bibinfo {pages} {793} (\bibinfo {year} {2003})}\BibitemShut
  {NoStop}%
\bibitem [{\citenamefont {Goldsmid}(2010)}]{Goldsmid:Thermoelectricity2010}%
  \BibitemOpen
  \bibfield  {author} {\bibinfo {author} {\bibfnamefont {H.~J.}\ \bibnamefont
  {Goldsmid}},\ }in\ \href@noop {} {\emph {\bibinfo {booktitle} {Introduction
  to Thermoelectricity}}},\ \bibinfo {series} {Springer Series in Materials
  Science}, Vol.~\bibinfo {volume} {1},\ \bibinfo {editor} {edited by\ \bibinfo
  {editor} {\bibfnamefont {R.}~\bibnamefont {Hull}}, \bibinfo {editor}
  {\bibfnamefont {J.}~\bibnamefont {Parisi}}, \bibinfo {editor} {\bibfnamefont
  {J.}~\bibnamefont {R.~M.~Osgood}}, \ and\ \bibinfo {editor} {\bibfnamefont
  {H.}~\bibnamefont {Warlimont}}}\ (\bibinfo  {publisher} {Springer},\ \bibinfo
  {year} {2010})\BibitemShut {NoStop}%
\bibitem [{\citenamefont {Guthrie}\ \emph {et~al.}(1949)\citenamefont
  {Guthrie}, \citenamefont {Wilson},\ and\ \citenamefont
  {Schomaker}}]{Guthrie:JChemPhys1949}%
  \BibitemOpen
  \bibfield  {author} {\bibinfo {author} {\bibfnamefont {G.}~\bibnamefont
  {Guthrie}}, \bibinfo {author} {\bibfnamefont {J.~N.}\ \bibnamefont {Wilson}},
  \ and\ \bibinfo {author} {\bibfnamefont {V.}~\bibnamefont {Schomaker}},\
  }\href@noop {} {\bibfield  {journal} {\bibinfo  {journal} {J.~Chem. Phys.}\
  }\textbf {\bibinfo {volume} {17}},\ \bibinfo {pages} {310} (\bibinfo {year}
  {1949})}\BibitemShut {NoStop}%
\bibitem [{\citenamefont {W\"{u}rger}(2010)}]{Wuerger:RepProgPhys2010}%
  \BibitemOpen
  \bibfield  {author} {\bibinfo {author} {\bibfnamefont {A.}~\bibnamefont
  {W\"{u}rger}},\ }\href@noop {} {\bibfield  {journal} {\bibinfo  {journal}
  {Rep. Prog. Phys.}\ }\textbf {\bibinfo {volume} {73}},\ \bibinfo {pages}
  {126601} (\bibinfo {year} {2010})}\BibitemShut {NoStop}%
\bibitem [{\citenamefont {Hills}\ \emph {et~al.}(1957)\citenamefont {Hills},
  \citenamefont {Jacobs},\ and\ \citenamefont
  {Lakshimi$\textrm{N}$arayanaiah}}]{Hill:Nature1957}%
  \BibitemOpen
  \bibfield  {author} {\bibinfo {author} {\bibfnamefont {G.~J.}\ \bibnamefont
  {Hills}}, \bibinfo {author} {\bibfnamefont {P.~W.~M.}\ \bibnamefont
  {Jacobs}}, \ and\ \bibinfo {author} {\bibfnamefont {N.}~\bibnamefont
  {Lakshimi$\textrm{N}$arayanaiah}},\ }\href@noop {} {\bibfield  {journal}
  {\bibinfo  {journal} {Nature}\ }\textbf {\bibinfo {volume} {179}},\ \bibinfo
  {pages} {96} (\bibinfo {year} {1957})}\BibitemShut {NoStop}%
\bibitem [{\citenamefont {Tasaka}\ \emph {et~al.}(1965)\citenamefont {Tasaka},
  \citenamefont {Morita},\ and\ \citenamefont
  {Nagasawa}}]{Tasaka:JChemPhys1965}%
  \BibitemOpen
  \bibfield  {author} {\bibinfo {author} {\bibfnamefont {M.}~\bibnamefont
  {Tasaka}}, \bibinfo {author} {\bibfnamefont {S.}~\bibnamefont {Morita}}, \
  and\ \bibinfo {author} {\bibfnamefont {M.}~\bibnamefont {Nagasawa}},\
  }\href@noop {} {\bibfield  {journal} {\bibinfo  {journal} {J.~Phys.~Chem.}\
  }\textbf {\bibinfo {volume} {69}},\ \bibinfo {pages} {4191} (\bibinfo {year}
  {1965})}\BibitemShut {NoStop}%
\bibitem [{\citenamefont {Tasaka}(1986)}]{Tasaka:PureApplChem1986}%
  \BibitemOpen
  \bibfield  {author} {\bibinfo {author} {\bibfnamefont {M.}~\bibnamefont
  {Tasaka}},\ }\href@noop {} {\bibfield  {journal} {\bibinfo  {journal} {Pure
  Appl. Chem.}\ }\textbf {\bibinfo {volume} {58}},\ \bibinfo {pages} {1637}
  (\bibinfo {year} {1986})}\BibitemShut {NoStop}%
\bibitem [{\citenamefont {Gaeta}\ \emph {et~al.}(1992)\citenamefont {Gaeta},
  \citenamefont {Ascolese}, \citenamefont {Bencivenga}, \citenamefont
  {de~Z\'arate}, \citenamefont {Pagliuca}, \citenamefont {Perna}, \citenamefont
  {Rossi},\ and\ \citenamefont {Mita}}]{Gaeta:JPhysChem1992}%
  \BibitemOpen
  \bibfield  {author} {\bibinfo {author} {\bibfnamefont {F.~S.}\ \bibnamefont
  {Gaeta}}, \bibinfo {author} {\bibfnamefont {E.}~\bibnamefont {Ascolese}},
  \bibinfo {author} {\bibfnamefont {U.}~\bibnamefont {Bencivenga}}, \bibinfo
  {author} {\bibfnamefont {J.~M.~O.}\ \bibnamefont {de~Z\'arate}}, \bibinfo
  {author} {\bibfnamefont {N.}~\bibnamefont {Pagliuca}}, \bibinfo {author}
  {\bibfnamefont {G.}~\bibnamefont {Perna}}, \bibinfo {author} {\bibfnamefont
  {S.}~\bibnamefont {Rossi}}, \ and\ \bibinfo {author} {\bibfnamefont {D.~G.}\
  \bibnamefont {Mita}},\ }\href@noop {} {\bibfield  {journal} {\bibinfo
  {journal} {J.~Phys. Chem.}\ }\textbf {\bibinfo {volume} {96}},\ \bibinfo
  {pages} {6342} (\bibinfo {year} {1992})}\BibitemShut {NoStop}%
\bibitem [{\citenamefont {Schoch}\ \emph {et~al.}(2008)\citenamefont {Schoch},
  \citenamefont {Han},\ and\ \citenamefont {Renaud}}]{Schoch:RevModPhys2008}%
  \BibitemOpen
  \bibfield  {author} {\bibinfo {author} {\bibfnamefont {R.~B.}\ \bibnamefont
  {Schoch}}, \bibinfo {author} {\bibfnamefont {J.}~\bibnamefont {Han}}, \ and\
  \bibinfo {author} {\bibfnamefont {P.}~\bibnamefont {Renaud}},\ }\href@noop {}
  {\bibfield  {journal} {\bibinfo  {journal} {Rev.~Mod.~Phys.}\ }\textbf
  {\bibinfo {volume} {80}},\ \bibinfo {pages} {839} (\bibinfo {year}
  {2008})}\BibitemShut {NoStop}%
\bibitem [{\citenamefont {McKemy}(2007)}]{McKemy:EurJPhysiol2007}%
  \BibitemOpen
  \bibfield  {author} {\bibinfo {author} {\bibfnamefont {D.~D.}\ \bibnamefont
  {McKemy}},\ }\href@noop {} {\bibfield  {journal} {\bibinfo  {journal}
  {Eur.~J.~Phy.}\ }\textbf {\bibinfo {volume} {454}},\ \bibinfo {pages} {777}
  (\bibinfo {year} {2007})}\BibitemShut {NoStop}%
\bibitem [{\citenamefont {Reid}\ and\ \citenamefont
  {Flonta}(2001)}]{Reid:Nature2001}%
  \BibitemOpen
  \bibfield  {author} {\bibinfo {author} {\bibfnamefont {G.}~\bibnamefont
  {Reid}}\ and\ \bibinfo {author} {\bibfnamefont {M.-L.}\ \bibnamefont
  {Flonta}},\ }\href@noop {} {\bibfield  {journal} {\bibinfo  {journal}
  {Nature}\ }\textbf {\bibinfo {volume} {413}},\ \bibinfo {pages} {480}
  (\bibinfo {year} {2001})}\BibitemShut {NoStop}%
\bibitem [{\citenamefont {Viana}\ \emph {et~al.}(2002)\citenamefont {Viana},
  \citenamefont {de~la Pe\~na},\ and\ \citenamefont
  {Belmonte}}]{Viana:NatureNeuro2002}%
  \BibitemOpen
  \bibfield  {author} {\bibinfo {author} {\bibfnamefont {F.}~\bibnamefont
  {Viana}}, \bibinfo {author} {\bibfnamefont {E.}~\bibnamefont {de~la Pe\~na}},
  \ and\ \bibinfo {author} {\bibfnamefont {C.}~\bibnamefont {Belmonte}},\
  }\href@noop {} {\bibfield  {journal} {\bibinfo  {journal} {Nat.~Neurosci.}\
  }\textbf {\bibinfo {volume} {5}},\ \bibinfo {pages} {254} (\bibinfo {year}
  {2002})}\BibitemShut {NoStop}%
\bibitem [{\citenamefont {Sandbakk}\ \emph {et~al.}(2013)\citenamefont
  {Sandbakk}, \citenamefont {Bentien},\ and\ \citenamefont
  {Kjelstrup}}]{Sandbakk:JMembrScie2013}%
  \BibitemOpen
  \bibfield  {author} {\bibinfo {author} {\bibfnamefont {K.~D.}\ \bibnamefont
  {Sandbakk}}, \bibinfo {author} {\bibfnamefont {A.}~\bibnamefont {Bentien}}, \
  and\ \bibinfo {author} {\bibfnamefont {S.}~\bibnamefont {Kjelstrup}},\
  }\href@noop {} {\bibfield  {journal} {\bibinfo  {journal} {J.~Membrane Sci.}\
  }\textbf {\bibinfo {volume} {434}},\ \bibinfo {pages} {10} (\bibinfo {year}
  {2013})}\BibitemShut {NoStop}%
\bibitem [{\citenamefont {Xu}\ \emph {et~al.}(2012)\citenamefont {Xu},
  \citenamefont {Liu}, \citenamefont {Lim}, \citenamefont {Qiao},\ and\
  \citenamefont {Chen}}]{Xu:NanoEnergy2012}%
  \BibitemOpen
  \bibfield  {author} {\bibinfo {author} {\bibfnamefont {B.}~\bibnamefont
  {Xu}}, \bibinfo {author} {\bibfnamefont {L.}~\bibnamefont {Liu}}, \bibinfo
  {author} {\bibfnamefont {H.}~\bibnamefont {Lim}}, \bibinfo {author}
  {\bibfnamefont {Y.}~\bibnamefont {Qiao}}, \ and\ \bibinfo {author}
  {\bibfnamefont {X.}~\bibnamefont {Chen}},\ }\href@noop {} {\bibfield
  {journal} {\bibinfo  {journal} {Nano Energy}\ }\textbf {\bibinfo {volume}
  {1}},\ \bibinfo {pages} {805–} (\bibinfo {year} {2012})}\BibitemShut
  {NoStop}%
\bibitem [{SM()}]{SM}%
  \BibitemOpen
  \href@noop {} {}\bibinfo {note} {See Supplemental Material [\textit{URL will
  be inserted by publisher}], which includes Refs.
  \cite{Derjaguin:NewYork1987,Lide:CRCPubComp2009,Takeyama:JPhysSocJap1983,Castellanos:Springer1998,
  Baldessari:JCollIntScie2008,Hiemstra:JCollIntScie1989,Leroy:JCollIntScie2013,Alekhin:KolloidZn1984,
  Oelkers:JSolChem1989,NISTWebbook2015,ChemSpider2015,Falasco:arXiv2014}.}\BibitemShut
  {Stop}%
\bibitem [{\citenamefont {Fair}\ and\ \citenamefont
  {Osterle}(1971)}]{Fair:JChemPhys1971}%
  \BibitemOpen
  \bibfield  {author} {\bibinfo {author} {\bibfnamefont {J.~C.}\ \bibnamefont
  {Fair}}\ and\ \bibinfo {author} {\bibfnamefont {J.~F.}\ \bibnamefont
  {Osterle}},\ }\href@noop {} {\bibfield  {journal} {\bibinfo  {journal}
  {J.~Chem.~Phys.}\ }\textbf {\bibinfo {volume} {54}},\ \bibinfo {pages} {3307}
  (\bibinfo {year} {1971})}\BibitemShut {NoStop}%
\bibitem [{\citenamefont {Sasidhar}\ and\ \citenamefont
  {Ruckenstein}(1982)}]{Sasidhar:JCollIntScie1982}%
  \BibitemOpen
  \bibfield  {author} {\bibinfo {author} {\bibfnamefont {V.}~\bibnamefont
  {Sasidhar}}\ and\ \bibinfo {author} {\bibfnamefont {E.}~\bibnamefont
  {Ruckenstein}},\ }\href@noop {} {\bibfield  {journal} {\bibinfo  {journal}
  {J.~Colloid.~Interface Sci.}\ }\textbf {\bibinfo {volume} {85}},\ \bibinfo
  {pages} {332} (\bibinfo {year} {1982})}\BibitemShut {NoStop}%
\bibitem [{\citenamefont {Snowdon}\ and\ \citenamefont
  {Turner}(1960{\natexlab{a}})}]{Snowdon:TransFaradaySoc1960II}%
  \BibitemOpen
  \bibfield  {author} {\bibinfo {author} {\bibfnamefont {P.~N.}\ \bibnamefont
  {Snowdon}}\ and\ \bibinfo {author} {\bibfnamefont {J.~C.~R.}\ \bibnamefont
  {Turner}},\ }\href@noop {} {\bibfield  {journal} {\bibinfo  {journal} {Trans.
  Faraday Soc.}\ }\textbf {\bibinfo {volume} {56}},\ \bibinfo {pages} {1812}
  (\bibinfo {year} {1960}{\natexlab{a}})}\BibitemShut {NoStop}%
\bibitem [{\citenamefont {Agar}\ and\ \citenamefont
  {Turner}(1960)}]{Agar:ProcRsSocLondA1960}%
  \BibitemOpen
  \bibfield  {author} {\bibinfo {author} {\bibfnamefont {J.~N.}\ \bibnamefont
  {Agar}}\ and\ \bibinfo {author} {\bibfnamefont {J.~C.~R.}\ \bibnamefont
  {Turner}},\ }\href@noop {} {\bibfield  {journal} {\bibinfo  {journal}
  {Proc.~R.~Soc.~A}\ }\textbf {\bibinfo {volume} {255}},\ \bibinfo {pages}
  {307} (\bibinfo {year} {1960})}\BibitemShut {NoStop}%
\bibitem [{\citenamefont {Snowdon}\ and\ \citenamefont
  {Turner}(1960{\natexlab{b}})}]{Snowdon:TransFaradaySoc1960}%
  \BibitemOpen
  \bibfield  {author} {\bibinfo {author} {\bibfnamefont {P.~N.}\ \bibnamefont
  {Snowdon}}\ and\ \bibinfo {author} {\bibfnamefont {J.~C.~R.}\ \bibnamefont
  {Turner}},\ }\href@noop {} {\bibfield  {journal} {\bibinfo  {journal} {Trans.
  Faraday Soc.}\ }\textbf {\bibinfo {volume} {56}},\ \bibinfo {pages} {1409}
  (\bibinfo {year} {1960}{\natexlab{b}})}\BibitemShut {NoStop}%
\bibitem [{\citenamefont {Buchner}\ \emph {et~al.}(1999)\citenamefont
  {Buchner}, \citenamefont {Hefter},\ and\ \citenamefont
  {May}}]{Buchner:PhysChemA1999}%
  \BibitemOpen
  \bibfield  {author} {\bibinfo {author} {\bibfnamefont {R.}~\bibnamefont
  {Buchner}}, \bibinfo {author} {\bibfnamefont {G.~T.}\ \bibnamefont {Hefter}},
  \ and\ \bibinfo {author} {\bibfnamefont {P.~M.}\ \bibnamefont {May}},\
  }\href@noop {} {\bibfield  {journal} {\bibinfo  {journal} {J.~
  Phys.~Chem.~A}\ }\textbf {\bibinfo {volume} {103}},\ \bibinfo {pages} {1}
  (\bibinfo {year} {1999})}\BibitemShut {NoStop}%
\bibitem [{\citenamefont {Tessier}\ and\ \citenamefont
  {Slater}(2006)}]{Tessier:Electrophoresis2006}%
  \BibitemOpen
  \bibfield  {author} {\bibinfo {author} {\bibfnamefont {F.}~\bibnamefont
  {Tessier}}\ and\ \bibinfo {author} {\bibfnamefont {G.~W.}\ \bibnamefont
  {Slater}},\ }\href@noop {} {\bibfield  {journal} {\bibinfo  {journal}
  {Electrophoresis}\ }\textbf {\bibinfo {volume} {27}},\ \bibinfo {pages} {686}
  (\bibinfo {year} {2006})},\ \bibinfo {note} {and references
  therein}\BibitemShut {NoStop}%
\bibitem [{\citenamefont {Bazant}\ \emph {et~al.}(2009)\citenamefont {Bazant},
  \citenamefont {Kilic}, \citenamefont {Storey},\ and\ \citenamefont
  {Ajdari}}]{Bazant:AdvCollIntScie2009}%
  \BibitemOpen
  \bibfield  {author} {\bibinfo {author} {\bibfnamefont {M.~Z.}\ \bibnamefont
  {Bazant}}, \bibinfo {author} {\bibfnamefont {M.~S.}\ \bibnamefont {Kilic}},
  \bibinfo {author} {\bibfnamefont {B.~D.}\ \bibnamefont {Storey}}, \ and\
  \bibinfo {author} {\bibfnamefont {A.}~\bibnamefont {Ajdari}},\ }\href@noop {}
  {\bibfield  {journal} {\bibinfo  {journal} {Adv.~Colloid Interface Sci.}\
  }\textbf {\bibinfo {volume} {152}},\ \bibinfo {pages} {48} (\bibinfo {year}
  {2009})},\ \bibinfo {note} {and references therein}\BibitemShut {NoStop}%
\bibitem [{\citenamefont {Cervera}\ \emph {et~al.}(2010)\citenamefont
  {Cervera}, \citenamefont {Ram\'irez}, \citenamefont {Mantanares},\ and\
  \citenamefont {Maf\'e}}]{Cervera:MicroNanofluidics2010}%
  \BibitemOpen
  \bibfield  {author} {\bibinfo {author} {\bibfnamefont {J.}~\bibnamefont
  {Cervera}}, \bibinfo {author} {\bibfnamefont {P.}~\bibnamefont {Ram\'irez}},
  \bibinfo {author} {\bibfnamefont {J.~A.}\ \bibnamefont {Mantanares}}, \ and\
  \bibinfo {author} {\bibfnamefont {S.}~\bibnamefont {Maf\'e}},\ }\href@noop {}
  {\bibfield  {journal} {\bibinfo  {journal} {Microfluid.~Nanofluid.}\ }\textbf
  {\bibinfo {volume} {9}},\ \bibinfo {pages} {41} (\bibinfo {year}
  {2010})}\BibitemShut {NoStop}%
\bibitem [{\citenamefont {Russel}\ \emph {et~al.}(1989)\citenamefont {Russel},
  \citenamefont {Saville},\ and\ \citenamefont
  {Schowalter}}]{Russel:CambrUnivPress1989}%
  \BibitemOpen
  \bibfield  {author} {\bibinfo {author} {\bibfnamefont {W.~B.}\ \bibnamefont
  {Russel}}, \bibinfo {author} {\bibfnamefont {D.~A.}\ \bibnamefont {Saville}},
  \ and\ \bibinfo {author} {\bibfnamefont {W.~R.}\ \bibnamefont {Schowalter}},\
  }in\ \href@noop {} {\emph {\bibinfo {booktitle} {Colloidal dispersions}}}\
  (\bibinfo  {publisher} {Cambridge University Press, New York},\ \bibinfo
  {year} {1989})\BibitemShut {NoStop}%
\bibitem [{\citenamefont {Movahed}\ and\ \citenamefont
  {Li}(2011)}]{Movahed:Electrophoresis2011}%
  \BibitemOpen
  \bibfield  {author} {\bibinfo {author} {\bibfnamefont {S.}~\bibnamefont
  {Movahed}}\ and\ \bibinfo {author} {\bibfnamefont {D.}~\bibnamefont {Li}},\
  }\href@noop {} {\bibfield  {journal} {\bibinfo  {journal} {Electrophoresis}\
  }\textbf {\bibinfo {volume} {32}},\ \bibinfo {pages} {1259} (\bibinfo {year}
  {2011})}\BibitemShut {NoStop}%
\bibitem [{\citenamefont {Daguji}\ \emph {et~al.}(2006)\citenamefont {Daguji},
  \citenamefont {Oka}, \citenamefont {Adachi},\ and\ \citenamefont
  {Shirano}}]{Daiguji:ElectrochemComm2006}%
  \BibitemOpen
  \bibfield  {author} {\bibinfo {author} {\bibfnamefont {H.}~\bibnamefont
  {Daguji}}, \bibinfo {author} {\bibfnamefont {Y.}~\bibnamefont {Oka}},
  \bibinfo {author} {\bibfnamefont {T.}~\bibnamefont {Adachi}}, \ and\ \bibinfo
  {author} {\bibfnamefont {K.}~\bibnamefont {Shirano}},\ }\href@noop {}
  {\bibfield  {journal} {\bibinfo  {journal} {Electrochem.~Comm.}\ }\textbf
  {\bibinfo {volume} {8}},\ \bibinfo {pages} {1796} (\bibinfo {year}
  {2006})}\BibitemShut {NoStop}%
\bibitem [{Com()}]{Comsol2014}%
  \BibitemOpen
  \href@noop {} {}\bibinfo {note} {Comsol Multiphysics, COMSOL
  \textsuperscript{\textregistered}, Inc., G\"ottingen, Germany,
  2014}\BibitemShut {NoStop}%
\bibitem [{\citenamefont {Behrens}\ and\ \citenamefont
  {Borkovec}(1999)}]{Behrens:PRE1999}%
  \BibitemOpen
  \bibfield  {author} {\bibinfo {author} {\bibfnamefont {S.~H.}\ \bibnamefont
  {Behrens}}\ and\ \bibinfo {author} {\bibfnamefont {M.}~\bibnamefont
  {Borkovec}},\ }\href@noop {} {\bibfield  {journal} {\bibinfo  {journal}
  {Phys.~Rev.~E}\ }\textbf {\bibinfo {volume} {60}},\ \bibinfo {pages} {7040}
  (\bibinfo {year} {1999})}\BibitemShut {NoStop}%
\bibitem [{\citenamefont {Reppert}\ and\ \citenamefont
  {Morgan}(2003)}]{Reppert:JGeoPhysRes2003}%
  \BibitemOpen
  \bibfield  {author} {\bibinfo {author} {\bibfnamefont {P.~M.}\ \bibnamefont
  {Reppert}}\ and\ \bibinfo {author} {\bibfnamefont {F.~D.}\ \bibnamefont
  {Morgan}},\ }\href@noop {} {\bibfield  {journal} {\bibinfo  {journal}
  {J.~Geophys. Res.}\ }\textbf {\bibinfo {volume} {108}},\ \bibinfo {pages}
  {2546} (\bibinfo {year} {2003})}\BibitemShut {NoStop}%
\bibitem [{\citenamefont {Behrens}\ and\ \citenamefont
  {Grier}(2001)}]{Behrens:JChemPhys2001}%
  \BibitemOpen
  \bibfield  {author} {\bibinfo {author} {\bibfnamefont {S.~H.}\ \bibnamefont
  {Behrens}}\ and\ \bibinfo {author} {\bibfnamefont {D.~G.}\ \bibnamefont
  {Grier}},\ }\href@noop {} {\bibfield  {journal} {\bibinfo  {journal}
  {J.~Chem.~Physics}\ }\textbf {\bibinfo {volume} {115}},\ \bibinfo {pages}
  {6716} (\bibinfo {year} {2001})}\BibitemShut {NoStop}%
\bibitem [{\citenamefont {Revil}\ and\ \citenamefont
  {Glover}(1997)}]{Revil:PhysRevB1997}%
  \BibitemOpen
  \bibfield  {author} {\bibinfo {author} {\bibfnamefont {A.}~\bibnamefont
  {Revil}}\ and\ \bibinfo {author} {\bibfnamefont {P.~W.~J.}\ \bibnamefont
  {Glover}},\ }\href@noop {} {\bibfield  {journal} {\bibinfo  {journal}
  {Phys.~Rev.~B}\ }\textbf {\bibinfo {volume} {55}},\ \bibinfo {pages} {1757}
  (\bibinfo {year} {1997})}\BibitemShut {NoStop}%
\bibitem [{\citenamefont {Einstein}(1905)}]{Einstein:AnnPhysik1905}%
  \BibitemOpen
  \bibfield  {author} {\bibinfo {author} {\bibfnamefont {A.}~\bibnamefont
  {Einstein}},\ }\href@noop {} {\bibfield  {journal} {\bibinfo  {journal}
  {Ann.~Phys.}\ }\textbf {\bibinfo {volume} {322}},\ \bibinfo {pages} {549–}
  (\bibinfo {year} {1905})}\BibitemShut {NoStop}%
\bibitem [{\citenamefont {Stolovitzky}(1998)}]{Stolovitzky:PhysicsLettA1998}%
  \BibitemOpen
  \bibfield  {author} {\bibinfo {author} {\bibfnamefont {G.}~\bibnamefont
  {Stolovitzky}},\ }\href@noop {} {\bibfield  {journal} {\bibinfo  {journal}
  {Phys.~Lett.~A}\ }\textbf {\bibinfo {volume} {241}},\ \bibinfo {pages} {240}
  (\bibinfo {year} {1998})}\BibitemShut {NoStop}%
\bibitem [{\citenamefont {Rogacs}\ and\ \citenamefont
  {Santiago}(2013)}]{Rogacs:AnalChem2013}%
  \BibitemOpen
  \bibfield  {author} {\bibinfo {author} {\bibfnamefont {A.}~\bibnamefont
  {Rogacs}}\ and\ \bibinfo {author} {\bibfnamefont {J.~G.}\ \bibnamefont
  {Santiago}},\ }\href@noop {} {\bibfield  {journal} {\bibinfo  {journal}
  {Anal.~Chem.}\ }\textbf {\bibinfo {volume} {85}},\ \bibinfo {pages} {5103}
  (\bibinfo {year} {2013})}\BibitemShut {NoStop}%
\bibitem [{\citenamefont {Braibanti}\ \emph {et~al.}(2008)\citenamefont
  {Braibanti}, \citenamefont {Vigolo},\ and\ \citenamefont
  {Piazza}}]{Braibanti:PRL2008}%
  \BibitemOpen
  \bibfield  {author} {\bibinfo {author} {\bibfnamefont {M.}~\bibnamefont
  {Braibanti}}, \bibinfo {author} {\bibfnamefont {D.}~\bibnamefont {Vigolo}}, \
  and\ \bibinfo {author} {\bibfnamefont {R.}~\bibnamefont {Piazza}},\
  }\href@noop {} {\bibfield  {journal} {\bibinfo  {journal} {Phys.~Rev.~Lett.}\
  }\textbf {\bibinfo {volume} {100}},\ \bibinfo {pages} {108303} (\bibinfo
  {year} {2008})}\BibitemShut {NoStop}%
\bibitem [{\citenamefont {Majee}\ and\ \citenamefont
  {W\"urger}(2011)}]{Majee:PRE2011}%
  \BibitemOpen
  \bibfield  {author} {\bibinfo {author} {\bibfnamefont {A.}~\bibnamefont
  {Majee}}\ and\ \bibinfo {author} {\bibfnamefont {A.}~\bibnamefont
  {W\"urger}},\ }\href@noop {} {\bibfield  {journal} {\bibinfo  {journal}
  {Phys.~Rev.~E.}\ }\textbf {\bibinfo {volume} {83}},\ \bibinfo {pages}
  {061403} (\bibinfo {year} {2011})}\BibitemShut {NoStop}%
\bibitem [{\citenamefont {Reichl}\ \emph {et~al.}(2014)\citenamefont {Reichl},
  \citenamefont {Herzog}, \citenamefont {G\"otz},\ and\ \citenamefont
  {Braun}}]{Reichl:PRL2014}%
  \BibitemOpen
  \bibfield  {author} {\bibinfo {author} {\bibfnamefont {M.}~\bibnamefont
  {Reichl}}, \bibinfo {author} {\bibfnamefont {M.}~\bibnamefont {Herzog}},
  \bibinfo {author} {\bibfnamefont {A.}~\bibnamefont {G\"otz}}, \ and\ \bibinfo
  {author} {\bibfnamefont {D.}~\bibnamefont {Braun}},\ }\href@noop {}
  {\bibfield  {journal} {\bibinfo  {journal} {Phys.~Rev.~Lett.}\ }\textbf
  {\bibinfo {volume} {112}},\ \bibinfo {pages} {198101} (\bibinfo {year}
  {2014})}\BibitemShut {NoStop}%
\bibitem [{\citenamefont {Derjaguin}\ \emph {et~al.}(1987)\citenamefont
  {Derjaguin}, \citenamefont {Churaev},\ and\ \citenamefont
  {Muller}}]{Derjaguin:NewYork1987}%
  \BibitemOpen
  \bibfield  {author} {\bibinfo {author} {\bibfnamefont {B.}~\bibnamefont
  {Derjaguin}}, \bibinfo {author} {\bibfnamefont {N.}~\bibnamefont {Churaev}},
  \ and\ \bibinfo {author} {\bibfnamefont {V.}~\bibnamefont {Muller}},\ }in\
  \href@noop {} {\emph {\bibinfo {booktitle} {Surface Forces}}}\ (\bibinfo
  {publisher} {Plenum, New York},\ \bibinfo {year} {1987})\BibitemShut
  {NoStop}%
\bibitem [{\citenamefont {Lide}(2009)}]{Lide:CRCPubComp2009}%
  \BibitemOpen
  \bibfield  {author} {\bibinfo {author} {\bibfnamefont {D.~R.}\ \bibnamefont
  {Lide}},\ }in\ \href@noop {} {\emph {\bibinfo {booktitle} {CRC Handbook of
  Chemistry and Physics}}},\ \bibinfo {editor} {edited by\ \bibinfo {editor}
  {\bibfnamefont {D.~R.}\ \bibnamefont {Lide}}}\ (\bibinfo  {publisher} {CRC
  Press, Boca Raton},\ \bibinfo {year} {2009})\BibitemShut {NoStop}%
\bibitem [{\citenamefont {Takeyama}\ and\ \citenamefont
  {Nakashima}(1983)}]{Takeyama:JPhysSocJap1983}%
  \BibitemOpen
  \bibfield  {author} {\bibinfo {author} {\bibfnamefont {N.}~\bibnamefont
  {Takeyama}}\ and\ \bibinfo {author} {\bibfnamefont {K.}~\bibnamefont
  {Nakashima}},\ }\href@noop {} {\bibfield  {journal} {\bibinfo  {journal}
  {J.~Phys.~Soc.~Jpn.}\ }\textbf {\bibinfo {volume} {52}},\ \bibinfo {pages}
  {2699} (\bibinfo {year} {1983})}\BibitemShut {NoStop}%
\bibitem [{\citenamefont {Castellanos}(1998)}]{Castellanos:Springer1998}%
  \BibitemOpen
  \bibfield  {author} {\bibinfo {author} {\bibfnamefont {A.}~\bibnamefont
  {Castellanos}},\ }in\ \href@noop {} {\emph {\bibinfo {booktitle}
  {Electrohydrodynamics}}},\ \bibinfo {editor} {edited by\ \bibinfo {editor}
  {\bibfnamefont {A.}~\bibnamefont {Castellanos}}}\ (\bibinfo  {publisher}
  {Springer, Wien},\ \bibinfo {year} {1998})\BibitemShut {NoStop}%
\bibitem [{\citenamefont {Baldessari}\ and\ \citenamefont
  {Santiago}(2008)}]{Baldessari:JCollIntScie2008}%
  \BibitemOpen
  \bibfield  {author} {\bibinfo {author} {\bibfnamefont {F.}~\bibnamefont
  {Baldessari}}\ and\ \bibinfo {author} {\bibfnamefont {J.~G.}\ \bibnamefont
  {Santiago}},\ }\href@noop {} {\bibfield  {journal} {\bibinfo  {journal}
  {J.~Colloid Interface Sci.}\ }\textbf {\bibinfo {volume} {325}},\ \bibinfo
  {pages} {526} (\bibinfo {year} {2008})}\BibitemShut {NoStop}%
\bibitem [{\citenamefont {Hiemstra}\ \emph {et~al.}(1989)\citenamefont
  {Hiemstra}, \citenamefont {Wit},\ and\ \citenamefont
  {Riemsdijk}}]{Hiemstra:JCollIntScie1989}%
  \BibitemOpen
  \bibfield  {author} {\bibinfo {author} {\bibfnamefont {T.}~\bibnamefont
  {Hiemstra}}, \bibinfo {author} {\bibfnamefont {J.~C. M.~D.}\ \bibnamefont
  {Wit}}, \ and\ \bibinfo {author} {\bibfnamefont {W.~H.~V.}\ \bibnamefont
  {Riemsdijk}},\ }\href@noop {} {\bibfield  {journal} {\bibinfo  {journal}
  {J.~Colloid Interface Sci.}\ }\textbf {\bibinfo {volume} {133}},\ \bibinfo
  {pages} {105} (\bibinfo {year} {1989})}\BibitemShut {NoStop}%
\bibitem [{\citenamefont {Leroy}\ \emph {et~al.}(2013)\citenamefont {Leroy},
  \citenamefont {Devau}, \citenamefont {Revil},\ and\ \citenamefont
  {Bizi}}]{Leroy:JCollIntScie2013}%
  \BibitemOpen
  \bibfield  {author} {\bibinfo {author} {\bibfnamefont {P.}~\bibnamefont
  {Leroy}}, \bibinfo {author} {\bibfnamefont {N.}~\bibnamefont {Devau}},
  \bibinfo {author} {\bibfnamefont {A.}~\bibnamefont {Revil}}, \ and\ \bibinfo
  {author} {\bibfnamefont {M.}~\bibnamefont {Bizi}},\ }\href@noop {} {\bibfield
   {journal} {\bibinfo  {journal} {J.~Colloid Interface Sci.}\ }\textbf
  {\bibinfo {volume} {410}},\ \bibinfo {pages} {81} (\bibinfo {year}
  {2013})}\BibitemShut {NoStop}%
\bibitem [{\citenamefont {Alekhin}\ \emph {et~al.}(1984)\citenamefont
  {Alekhin}, \citenamefont {Sidorova}, \citenamefont {Ivanova},\ and\
  \citenamefont {Lakshtanov}}]{Alekhin:KolloidZn1984}%
  \BibitemOpen
  \bibfield  {author} {\bibinfo {author} {\bibfnamefont {Y.~V.}\ \bibnamefont
  {Alekhin}}, \bibinfo {author} {\bibfnamefont {M.~P.}\ \bibnamefont
  {Sidorova}}, \bibinfo {author} {\bibfnamefont {L.~I.}\ \bibnamefont
  {Ivanova}}, \ and\ \bibinfo {author} {\bibfnamefont {L.~Z.}\ \bibnamefont
  {Lakshtanov}},\ }\href@noop {} {\bibfield  {journal} {\bibinfo  {journal}
  {Colloid J.~USSR.}\ }\textbf {\bibinfo {volume} {46}},\ \bibinfo {pages}
  {1032} (\bibinfo {year} {1984})}\BibitemShut {NoStop}%
\bibitem [{\citenamefont {Oelkers}\ and\ \citenamefont
  {Helgeson}(1989)}]{Oelkers:JSolChem1989}%
  \BibitemOpen
  \bibfield  {author} {\bibinfo {author} {\bibfnamefont {E.~H.}\ \bibnamefont
  {Oelkers}}\ and\ \bibinfo {author} {\bibfnamefont {H.~C.}\ \bibnamefont
  {Helgeson}},\ }\href@noop {} {\bibfield  {journal} {\bibinfo  {journal}
  {J.~Solution Chem.}\ }\textbf {\bibinfo {volume} {18}},\ \bibinfo {pages}
  {601} (\bibinfo {year} {1989})}\BibitemShut {NoStop}%
\bibitem [{NIS()}]{NISTWebbook2015}%
  \BibitemOpen
  \href@noop {} {}\bibinfo {note} {National Institute of Standards and
  Technology, http://webbook.nist.gov}\BibitemShut {NoStop}%
\bibitem [{Che()}]{ChemSpider2015}%
  \BibitemOpen
  \href@noop {} {}\bibinfo {note}
  {Royal\:Society\:of\:Chemistry,\:http://www.chemspider.com}\BibitemShut
  {NoStop}%
\bibitem [{\citenamefont {Falasco}\ \emph {et~al.}()\citenamefont {Falasco},
  \citenamefont {Gnann},\ and\ \citenamefont {Kroy}}]{Falasco:arXiv2014}%
  \BibitemOpen
  \bibfield  {author} {\bibinfo {author} {\bibfnamefont {G.}~\bibnamefont
  {Falasco}}, \bibinfo {author} {\bibfnamefont {M.~V.}\ \bibnamefont {Gnann}},
  \ and\ \bibinfo {author} {\bibfnamefont {K.}~\bibnamefont {Kroy}},\
  }\href@noop {} {\bibinfo  {journal} {arXiv:1406.2116v1}\ }\BibitemShut
  {NoStop}%
\end{thebibliography}%

\end{document}